\documentstyle[prl,aps]{revtex}
\twocolumn
\begin{document}
\input{epsf}
\draft
\twocolumn[\hsize\textwidth\columnwidth\hsize\csname@twocolumnfalse\endcsname
\title{Parametric autoresonance}
\author{Evgeniy Khain and Baruch Meerson  }
\address{Racah Institute  of  Physics, Hebrew
University   of  Jerusalem, Jerusalem 91904, Israel} \maketitle

\begin{abstract}
We investigate parametric
autoresonance: a persisting phase locking which
occurs
when the driving frequency of a parametrically excited nonlinear
oscillator
slowly varies  with
time. In this regime,
the resonant excitation is continuous and unarrested
by the oscillator nonlinearity. The
system
has three
characteristic time scales, the fastest one corresponding to
the natural frequency of
the oscillator. We perform averaging over the
fastest time scale and analyze the reduced set of equations
analytically and numerically. Analytical results are obtained
by exploiting the scale separation between the two remaining time
scales which enables one to use the
adiabatic invariant of the perturbed nonlinear motion.
\end{abstract}
 \vskip1pc] \narrowtext \pagebreak

\section{INTRODUCTION}

This work addresses a combined action of two
mechanisms of resonant excitation of (classical) nonlinear oscillating
systems. The first is {\it parametric resonance}. The
second is {\it autoresonance}.

There are numerous oscillatory systems which interaction with the
external world amounts only to a periodic time dependence of their
parameters. The corresponding resonance is called {\it parametric}
\cite{Landau,Bogolubov}. A textbook
example is a simple pendulum with a vertically
oscillating point of suspension \cite{Landau}. The main resonance
occurs when the excitation frequency
$\omega$ is nearly twice the natural frequency of the oscillator
$\omega_{0}$ \cite{Landau,Bogolubov}. Applications of this basic
phenomenon in physics and technology are ubiquitous.

{\it Autoresonance} occurs in nonlinear oscillators
driven by a small {\it external} force, almost periodic in time.
If the force is
{\it exactly}
periodic, and in resonance with the natural frequency of the oscillator,
the resonance region  of the phase plane
has a finite (and relatively small) width \cite{Sagdeev,Lichtenberg}.
If instead the driving
frequency is slowly varying in time (in the right direction
determined by the nonlinearity sign), the oscillator can stay
phase-locked despite the nonlinearity. This leads to a continuous
resonant excitation. Autoresonance has found
many applications. It was extensively
studied
in the context of relativistic particle acceleration: in
the 40-ies
by McMillan \cite{McMillan}, Veksler \cite{Veksler} and
Bohm and Foldy \cite{Bohm1,Bohm2}, and more recently
\cite{Golovanivsky2,Meerson2,Meerson10,Friedland1}.
Additional applications include a quasiclassical scheme
of excitation of atoms \cite{Meerson1}
and molecules \cite{molecules}, excitation of nonlinear
waves \cite{Deutsch,Friedland2},
solitons \cite{Aranson,Friedland3},
vortices \cite{Friedland4,Friedland5} and other collective
modes \cite{Fajans1}
in fluids and
plasmas,
an autoresonant mechanism of transition to chaos in Hamiltonian systems
\cite{Yariv1,Cohen}, etc.

Until now autoresonance was considered only in systems
executing {\it externally} driven oscillations. In this work we
investigate autoresonance in a {\it parametrically} driven
oscillator.

Our presentation will be as follows. In Section 2 we briefly review
the parametric resonance in non-linear oscillating systems.
Section 3 deals, analytically and numerically,
with parametric autoresonance. The conclusions are presented
in Section 4. Some details of derivation are
given in Appendices A and B.

\section{PARAMETRIC RESONANCE WITH A CONSTANT DRIVING FREQUENCY}

The parametric resonance
in a weakly nonlinear oscillator with finite dissipation and
detuning is describable by the
following equation of motion \cite{Bogolubov,Struble2,Morozov}:
\begin{equation}
\ddot{x}+2\gamma\dot{x}+
\left[1+\epsilon \cos \left((2+\delta)t\right)\right]x- \beta
x^3=0. \label{Bbegin1}
\end{equation}
where the units of time are chosen in such a way that the scaled
natural frequency of the oscillator in the small-amplitude limit
is equal to 1. In Eq. (\ref{Bbegin1}) $ \epsilon$ is the amplitude
of the driving force, which is assumed to be small: $
0<\epsilon\ll1$,  $ \delta \ll 1$ is the detuning parameter, $
\gamma$ is the (scaled) damping coefficient $ ( 0<\gamma\ll1 ) $
and $\beta$ is the nonlinearity coefficient. For concreteness we
assume $\beta>0$ (for a pendulum $\beta =1/6$).

Working in
the limit of weak nonlinearity, dissipation and driving, we can
employ the method of
averaging \cite{Bogolubov,Sagdeev,Rabinovich,Drazin},
valid for most of the initial conditions \cite{Sagdeev,Lichtenberg}.
The unperturbed oscillation period is the fast time.
Putting $ x=a(t)\cos\theta(t)$ and $
\dot{x}=-a(t)\sin\theta(t)$ and performing averaging over the fast
time, we
arrive at the averaged equations
\begin{eqnarray}
\dot{a}&=&-\gamma a+\frac{\epsilon a}{4}\sin 2\psi,\nonumber
\\
\dot{\psi}&=& -\frac{\delta}{2}-\frac{3\beta a^2}{8}
+\frac{\epsilon}{4}\cos 2\psi, \label{B6}
\end{eqnarray}
where a new phase $\psi=\theta-[(2+\delta)/2] t$ has been
introduced. The averaged
system  (\ref{B6}) is an autonomous dynamical
system with two degree of freedom
and therefore integrable.
In the conservative case  $\gamma=0$ Eqs. (\ref{B6})
become:
\begin{eqnarray}
\dot{a}&=&\frac{\epsilon a}{4}\sin 2\psi,\nonumber
\\
\dot{\psi}&=&-\frac{\delta}{2}-\frac{3\beta
a^2}{8}+\frac{\epsilon}{4}\cos 2\psi.\label{B61}
\end{eqnarray}
As $\sin 2\psi$ and $\cos 2\psi$ are periodic functions of $\psi $
with a period $\pi$, it is sufficient to consider the interval
$-\pi/2<\psi\leq\pi/2 $. For small enough detuning, $
\delta<\epsilon/2 $, there is an elliptic fixed point with a
non-zero
amplitude:$$a_{*}=\pm\left[\frac{2\epsilon}{3\beta}\left(1-\frac{2\delta}{\epsilon}\right)\right]^{1/2};
\indent\psi_{*}=0.$$ We need to calculate the period of motion in
the phase plane along a closed orbit around this fixed point (such
an orbit is shown in Fig.~\ref{close}).

\begin{figure}[h]
\vspace{-0.3 cm} \center{\epsfxsize=7.5 cm 
\epsffile{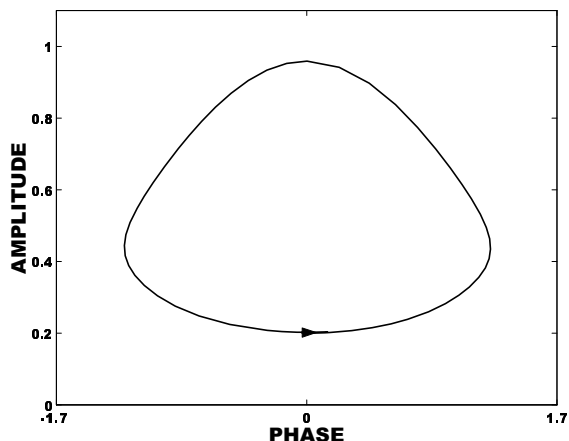}} \caption{Parametric resonance with a
constant driving frequency. Shown is a typical closed orbit in the
phase plane with period $T$ ($\epsilon=0.04, \delta=-0.04,
\beta=1/6$ and $\gamma=0.$). For a time-dependent driving
frequency $\nu(t)$, the autoresonance will occur if the
characteristic time for variation of $\nu(t)$ is much greater than
$T$, see criterion~(\ref{Ccrit}).  \label{close}}
\end{figure}

This calculation was performed
by Struble \cite{Struble2}.
For a zero detuning,
$\delta=0$,
Hamilton's function (we will call it the Hamiltonian) of
the system~(\ref{B61})
is the following:

\begin{equation}
H(I,\psi)=\frac{\epsilon I}{4}\cos 2\psi-\frac{3\beta
I^2}{8}=H_{0} =  const., \label{help1}
\end{equation}
where we have introduced the action variable
$I=a^2/2$. Solving Eq.~(\ref{help1})
for $I$ and substituting the result into the
Hamilton's equation for $\dot{\psi}$ we obtain:
\begin{equation}
\dot{\psi}=\mp\frac{\epsilon}{4}\left(\cos^2 2\psi-\frac{24\beta
H_{0}}{\epsilon^2}\right)^{1/2}, \label{help2}
\end{equation}
where the minus (plus) sign corresponds to the upper (lower) part of the closed
orbit.  The
period of the amplitude and phase oscillations is therefore
\begin{equation}
T=\frac{8}{\epsilon}\int_{-\overline{\psi}}^{\overline{\psi}}\frac{d\psi}{\left(\cos^2
2\psi-\frac{24\beta H_{0}}{\epsilon^2}\right)^{1/2}},
\label{help3}
\end{equation}
where $-\overline{\psi}$ and $\overline{\psi}$ are the roots of
the equation $ \cos^2 2\psi=24\beta H_{0}/\epsilon^2.$ Calculating the
integral, we
obtain
\begin{equation}
T=\frac{8}{\epsilon}K(m), \label{help4}
\end{equation}
where $K(m)$ is the complete
elliptic integral of the first kind \cite{Abramowitz},
and $m=1-24\beta H_{0}/\epsilon^2$. This
result will be used in Section 3 to establish a necessary
condition for the parametric autoresonance to
occur.

\section{PARAMETRIC RESONANCE WITH A TIME-DEPENDENT DRIVING
FREQUENCY: PARAMETRIC AUTORESONANCE}

Now let the driving frequency
vary with time. This time dependence introduces
an additional (third) time scale into the problem.
The governing equation becomes
\begin{equation}
\ddot{x}+ 2\gamma\dot{x} + (1+\epsilon\cos\phi)x-\beta x^3=0,
\label{B6.5}
\end{equation}
where $\dot{\phi}=\nu(t)$. We will assume $\nu(t)$ to be a
{\it slowly} decreasing function which initial value
is $\nu(t=0)=2+\delta.$  Using the scale separation, we obtain
the averaged equations.  The averaging
procedure of Section 2 can be repeated by
replacing $(2+\delta)t$ by $\phi$ in all equations.
There is one new point that
should be treated more accurately.
The averaging procedure is applicable (again, for most
of the initial conditions) if there is a separation
of time
scales. It requires, in particular,
a strong inequality $2\dot{\theta}+\nu (t)\gg 2\dot{\theta}-\nu(t)$.
This inequality can limit the time of validity of the method of
averaging.  Let us assume, for concreteness, a linear frequency ``chirp'':
\begin{equation}
\nu(t)=2+\delta-2\mu t, \label{B7}
\end{equation}
where $\mu\ll 1$ is the
chirp rate. In this case the averaging procedure is
valid as long as $\mu t\ll 1$.

Introducing a new phase $\psi=\theta-\phi/2 $,
we obtain a reduced set of equations (compare to
Eqs.~(\ref{B6})):
\begin{eqnarray}
\dot{a}&=&-\gamma a+\frac{\epsilon a}{4}\sin 2\psi,\nonumber
\\
\dot{\psi}&=&-\frac{\delta}{2}+\mu t-\frac{3\beta
a^2}{8}+\frac{\epsilon}{4}\cos 2\psi. \label{B8}
\end{eqnarray}
The first of Eqs. (\ref{B8})
is typical for {\it  parametric} resonance: to get excitation one should
start from a non-zero oscillation amplitude. As we will see,
the $\mu t$ term in the second of Eqs. (\ref{B8}) (when small enough
and
of the right sign) provides a continuous phase locking, similar to
the externally driven
autoresonance.

\begin{figure}[h]
\vspace{-0.4 cm} \center{\epsfxsize=7.5 cm 
\epsffile{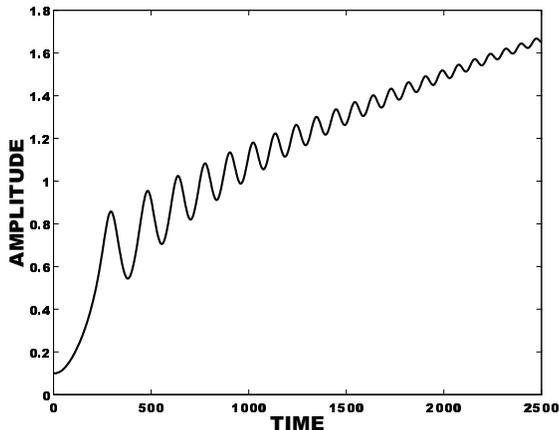}}

\caption{An example of parametric autoresonance. Shown is the
oscillation amplitude versus time, computed numerically from the averaged
equations~(\ref{B8}). The system remains phase-locked
which allows the amplitude to increase, on the average,
with time. The parameters are
$\mu=6.5\cdot 10^{-5}, \epsilon=0.04, \delta=-0.01,
\beta=1/6$ and $\gamma=0.001$. \label{figexample}}
\end{figure}

Consider a numerical example. Fig.~\ref{figexample} shows
the time dependence $a (t)$ found by solving
Eqs. ~(\ref{B8}) numerically. One
can see that the system remains
phase locked  which allows the amplitude of oscillations
to increase, on the average, with time in spite of the nonlinearity.
The time-dependence of the amplitude includes
a slow trend and relatively
fast, decaying oscillations. These are the two time
scales
remaining after the averaging over the fastest time scale.

Similar to the externally-driven autoresonance, a persistent
growth of the oscillation amplitude requires the characteristic
time of variation of $\nu(t)$ to be much greater than the
``nonlinear''  period $T$ [see Eq.~(\ref{help4})] of oscillations
of the amplitude:
\begin{equation}
\left|\frac{\nu(t)}{\dot{\nu}(t)}\right|\gg T. \label{Ccrit}
\end{equation}

Like its externally-driven analog, the
parametric autoresonance is insensitive to the exact form of
$\nu(t)$.  For a given set of
parameters, the optimal chirping rate can be found: too low a
chirping rate means an inefficient excitation, while too high a rate
leads to phase unlocking and termination of the excitation.

In the remainder of the paper we will develop an analytical theory
of the parametric autoresonance. The first objective of this
theory is a description of
the slow trend in the amplitude (and phase) dynamics.
When the driving frequency $\nu$ is constant, there is an elliptic
fixed point $a_{*}$ (see Section 2). When $\nu$ varies with time,
the fixed point ceases to exist. However, for a {\it slowly}-varying
$\nu(t)$ one can define a ``quasi-fixed'' point $a_{*}(t)$ which
is a slowly varying function of time. It is this quasi-fixed point
that represents the slow trend seen in Fig. \ref{figexample} and
corresponds to an ``ideal'' phase-locking regime. The
fast, decaying oscillations seen in Fig. \ref{figexample}
correspond to oscillations around the quasi-fixed point in the
phase plane [this phase plane is actually projection of the
extended phase space ($a,\psi, t$) on the ($a,\psi$)-plane].

In the main part of this Section we neglect the dissipation
and use a Hamiltonian
formalism. First we will consider excitation in
the vicinity of the quasi-fixed point. Then
excitation from arbitrary initial conditions will be investigated.
Finally, the role of dissipation will be briefly analyzed.

For a time-dependent $\nu(t)$,  the
Hamiltonian  becomes [compare to Eq. (\ref{help1})]:
\begin{equation}
H(I,\psi,t)=\frac{\epsilon I}{4}\left(\alpha(t)+\cos
2\psi\right)-\frac{3\beta I^2}{8}, \label{C1}
\end{equation}
where $\alpha(t)=(4/\epsilon)(1-\nu(t)/2).$ The
Hamilton's equations are:
\begin{eqnarray}
\dot{I}&=&\frac{\epsilon I}{2}\sin 2\psi,\nonumber
\\
\dot{\psi}&=&\frac{\epsilon}{4}\left(\alpha+\cos
2\psi\right)-\frac{3\beta I}{4}. \label{C2}
\end{eqnarray}
Let us find the quasi-fixed point of~(\ref{C2}), i.e.
the special autoresonance trajectory $I_{*}(t),\,\psi_{*}(t)$
corresponding to
the ``ideal'' phase locking (a pure trend without oscillations).

Assuming a slow time dependence, we put $\dot{\psi_{*}} = 0$,
that is
\begin{equation}
\frac{\epsilon}{4}\left(\alpha+\cos 2\psi_{*}\right)-\frac{3\beta
I_{*}}{4}=0. \label{C2-1}
\end{equation}
Differentiating it with respect to time and using Eqs.
(\ref{C2}), we obtain an algebraic equation for $\psi_* (t)$:
\begin{equation}
2\alpha (t) \sin 2\psi_{*}+\sin 4\psi_{*} = \frac{16\mu}{\epsilon^2} .
\label{C2-2}
\end{equation}
At this point we should demand that $\dot{\psi_*} (t)$,
evaluated
on the solution of Eq. (\ref{C2-2}),
is indeed negligible compared to the rest of terms in the
equation (\ref{C2}) for $\dot{\psi} (t)$. It is easy to see that
this requires $16 \mu/\epsilon^2 \ll 1$.
In this case the sines in Eq. (\ref{C2-2}) can be
replaced by their arguments, and we obtain
the following simple expressions for the
quasi-fixed point:
\begin{eqnarray}
I_{*}&\simeq&\frac{\epsilon}{3\beta}\left(\alpha+1\right)\,,\nonumber
\\
\psi_{*}&\simeq&\frac{k}{\alpha+1}\,, \label{C3}
\end{eqnarray}
where $k=4\mu/\epsilon^2$.

\subsection{Excitation in the vicinity of the quasi-fixed point}

\begin{figure}[h]
\vspace{0 cm} \rightline{\epsfxsize=7.5 cm 
\epsffile{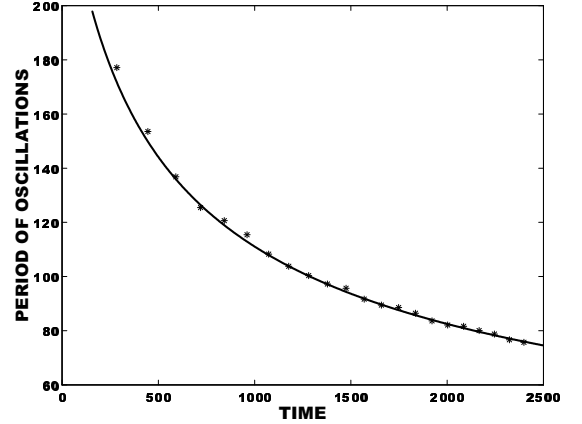}}

\caption{Excitation in the vicinity of the quasi-fixed point:
the time-dependence of the period $T$ of the action and phase
oscillations. The solid line is the theoretical curve, Eq.~(\ref{C10}),
the asterisks are points obtained numerically. The parameters are
$\mu=6.5\cdot 10^{-5}, \epsilon=0.04, \delta=-0.01$ and
$\beta=1/6$. \label{fiq-C3}}
\end{figure}

Let us make the
canonical transformation from
variables $I$ and $\psi$ to $ \delta I=I-I_{*}$ and
$\delta\psi=\psi-\psi_{*}.$ Assuming $\delta I$ and $\delta\psi$
to be small and keeping terms up to
the second order in $ \delta I$ and $\delta\psi$,
we obtain the new Hamiltonian:
\begin{eqnarray}
H(\delta I,\delta\psi,\alpha(t))&=&
-\frac{\epsilon k}{\alpha+1}\delta I \delta\psi- \nonumber
\\
&-&\frac{3\beta}{8}(\delta I)^2-
\frac{\epsilon^2}{6\beta}(\alpha+1)(\delta\psi)^2\,. \label{C4}
\end{eqnarray}
Here and in the following small terms of order of $k^2$ are neglected.
Let us start with the calculation of the local maxima of $\delta I
(t)$
and $\delta\psi (t)$, which will be called
$\delta I_{max}(t)$ and $\delta\psi_{max}(t)$, respectively.
As $\alpha(t)$ is a
slow function of time [so that
the strong inequality (\ref{Ccrit}) is satisfied], we can
exploit the approximate
constancy of the adiabatic invariant \cite{Landau,Goldstein}:
\begin{equation}
J=\frac{1}{2\pi}\oint\delta I d(\delta\psi) \simeq const. \label{C5}
\end{equation}
$|J|$ is the
area of the ellipse defined by Eq.~(\ref{C4}) with the
time-dependencies ``frozen''. Therefore,
\begin{equation}
J=\frac{2}{\epsilon}\frac{H}{(\alpha+1)^{1/2}}
\simeq const. \label{C6}
\end{equation}
This expression can be rewritten in terms of $\delta
I$ and $\delta\psi$:
\begin{eqnarray}
|J| &=&\frac{2k}{(\alpha+1)^{3/2}}\delta I
\delta\psi+\frac{3\beta}{4\epsilon}\frac{1}{(\alpha+1)^{1/2}}(\delta
I)^2\nonumber
\\
&+&\frac{\epsilon}{3\beta}(\alpha+1)^{1/2}(\delta\psi)^2.
\label{C7}
\end{eqnarray}
If $k=4 \mu/\epsilon^2 \ll 1$, the term with $\delta I
\delta\psi$ in (\ref{C7}) can be neglected (in this approximation
one has $\psi_* = 0$).
Then $J$ becomes a sum of two non-negative terms, one of
them having the maximum value when the other one vanishes. Therefore,
\begin{equation}
\delta I_{max}(t)=2\left(\frac{\epsilon
J}{3\beta}\right)^{1/2}\left(\alpha+1\right)^{1/4}, \label{C8}
\end{equation}
and
\begin{equation}
\delta\psi_{max}(t)=\left(\frac{3\beta
J}{\epsilon}\right)^{1/2}\frac{1}{(\alpha+1)^{1/4}}. \label{C9}
\end{equation}
Now we calculate the period of oscillations of the action and
phase. Using the well-known relation \cite{Landau}
$T=2\pi(\partial J/\partial H)$, we obtain from Eq. (\ref{C6}):
\begin{equation}
T=\frac{4\pi}{\epsilon}\frac{1}{(\alpha+1)^{1/2}}. \label{C10}
\end{equation}
The period of oscillations versus time is shown in
Fig.~\ref{fiq-C3}. The theoretical curve [Eq.~(\ref{C10})] shows
an excellent agreement with the numerical solution.

Now we obtain the complete solution $\delta I(t)$ and
$\delta\psi(t)$. The Hamilton's equations corresponding to the
Hamiltonian (\ref{C4}) are:

\begin{eqnarray}
\dot{\delta I}&=&
\frac{\epsilon^2}{3\beta}\left(\alpha+1\right)\delta\psi+\frac{\epsilon
k}{\alpha+1}\delta I,\nonumber
\\
\dot{\delta\psi}&=&-\frac{3\beta}{4}\delta I-\frac{\epsilon
k}{\alpha+1}\delta\psi. \label{C11}
\end{eqnarray}

Differentiating the second equation with respect to time and
substituting the first one, we obtain a linear differential
equation for $\delta\psi(t)$:
\begin{equation}
\ddot{\delta\psi}+\omega^2(t)\delta\psi=0, \label{C12}
\end{equation}
where $\omega(t)=(\epsilon/2)(\alpha(t)+1)^{1/2}$. For the
linear $\nu(t)$ dependence  (Eq.~(\ref{B7})) we
have $\alpha(t)=4\mu t/\epsilon-2\delta/\epsilon$, therefore for
$k\ll 1$ the criterion $\dot{\omega}/\omega^2\ll 1$ is satisfied,
and Eq. (\ref{C12}) can be solved by the WKB method (see, {\it
e.g.} \cite{Lichtenberg}).

\begin{figure}[h]
\vspace{0 cm} \leftline{\epsfxsize=7.5 cm 
\epsffile{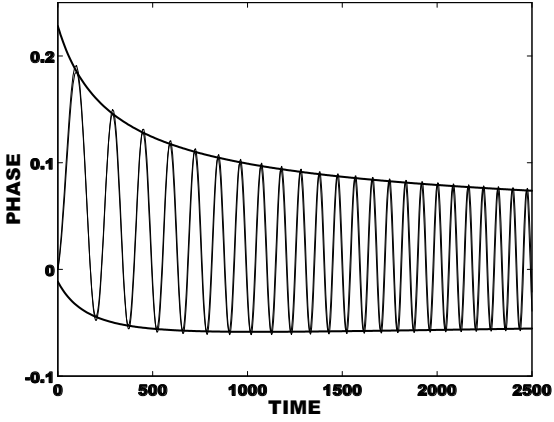}} \caption{Parametric autoresonance
excitation in the vicinity of the quasi-fixed point. Shown is the
phase $\psi (t)$ found analytically [Eqs. (\ref{C3}) and (\ref{C15})]
and by solving Eq. (\ref{C2}) numerically. The
analytical and numerical curves are indistinguishable. Also shown
are the minimum and maximum phase deviations predicted by Eq.
(\ref{C9}) and (\ref{C3}). The parameters are the same as in Fig.
\ref{fiq-C3}. \label{fig-c2}}
\end{figure}

The WKB solution takes the form (details are given in Appendix A):
\begin{eqnarray}
\delta\psi(t)&=&\left(\frac{3\beta
J}{\epsilon}\right)^{1/2}\frac{1}{(\alpha+1)^{1/4}}\nonumber
\\
& \times & \cos\left(q_{0}+\frac{(\alpha+1)^{3/2}}{3k}\right),
\label{C15}
\end{eqnarray}
where the phase $q_{0}$ is determined by the initial conditions.
The full solution for the phase is $\psi=\delta\psi+\psi_{*}$ and Fig.
~\ref{fig-c2} compares it with a numerical solution of Eq.
(\ref{C2}). Also shown are the minimum and maximum
phase deviations predicted by Eqs. (\ref{C9}) and (\ref{C3}). One
can see that the agreement is excellent.

The solution for $\delta I (t)$ can be obtained by substituting
Eq.~(\ref{C15}) into the second equation of the
system~(\ref{C11}). In the same order of accuracy (see Appendix A)
\begin{equation}
\delta I(t)=2\left(\frac{\epsilon
J}{3\beta}\right)^{1/2}\left(\alpha+1\right)^{1/4}\sin\left(q_{0}+\frac{(\alpha+1)^{3/2}}{3k}\right)\,.
\label{C19}
\end{equation}

\begin{figure}[h]
\vspace{0 cm} \leftline{\epsfxsize=7.5 cm 
\epsffile{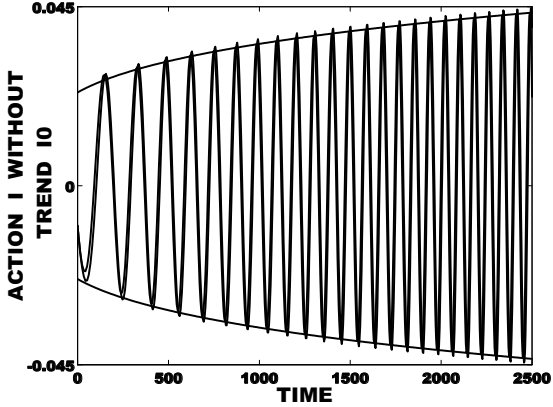}} \caption{Parametric autoresonance
excitation in the vicinity of
the quasi-fixed point. Shown is the action variable
$\delta I (t)$ from Eq.~(\ref{C19}) and from the numerical solution.
Also shown are the minimum and maximum action
deviations predicted by Eq. (\ref{C8}). The parameters are the same as
in Fig. \ref{fiq-C3}.
\label{fig-c1}}
\end{figure}

Fig.~\ref{fig-c1} shows the dependence of the action variable with
the trend $I_*(t)$ subtracted, $\delta I (t)$, on time predicted
by Eq.~(\ref{C19}), and found from the numerical solution. It also
shows the minimum and maximum action deviations (\ref{C8}). Again,
a very good agreement is obtained.

\subsection{Excitation from arbitrary initial conditions}

In this Subsection we go beyond
the close vicinity of the quasi-fixed
point
and
calculate the maximum
deviations of the action $I$ and phase $\psi$
for arbitrary initial conditions. Again, these calculations are made
possible by employing the adiabatic invariant for the general case.
Correspondingly,
the period of the action and phase oscillations will
be also calculated.

Let us first express the maximum and minimum {\it action} deviations
in terms of the Hamiltonian $H$ and driving frequency
$\nu(t)$. Solving Eq.~(\ref{C1}) as a quadratic equation for $I$,
we obtain: $$ I_{1,2}=\frac{\epsilon}{3\beta}\left(\alpha+\cos
2\psi\right)\pm\left[ \frac{\epsilon^2}{9\beta^2}\left(\alpha+\cos
2\psi\right)^2-\frac{8H}{3\beta}\right]^{1/2}.$$ The time derivative
of
$I$ vanishes when $I=I_{max}$
or $I=I_{min}$. Therefore, from
the first equation of the system~(\ref{C2}) $\psi=0$ so that
\begin{equation}
I_{max,min}=\frac{\epsilon}{3\beta}\left(\alpha+1\right)\pm\left[
\frac{\epsilon^2}{9\beta^2}\left(\alpha+1\right)^2-\frac{8H_{up,down}}{3\beta}\right]^{1/2},
\label{C20}
\end{equation}
where $H_{up, down}=H(I_{max, min},\psi=0)$.

Now we express the maximum and minimum {\it phase} deviations
through the Hamiltonian $H$ and driving frequency $\nu(t)$. The
time derivative $\dot{\psi}$ vanishes if $\psi=\psi_{max}$ or
$\psi=\psi_{min}$, then the second equation of the system
~(\ref{C2}) yields $I=(\epsilon/3\beta)(\alpha+\cos 2\psi)$. In
this case the Hamiltonian~(\ref{C1}) becomes
$H_{right,left}=(\epsilon^2/24\beta)(\alpha+\cos
2\psi_{max,min})^2$. Finally, the expression for $\psi_{max,min}$
is
\begin{equation}
\psi_{max,min}=\pm\frac{1}{2}\arccos\left[\left(\frac{24\beta
H_{right,left}}{\epsilon^2}\right)^{1/2}-\alpha\right]\,. \label{C21}
\end{equation}

\begin{figure}[h]
\vspace{-0.1 cm} \rightline{\epsfxsize=7.5 cm 
\epsffile{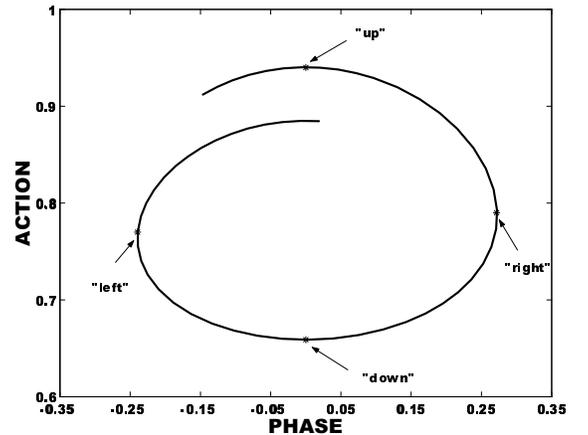}} \caption{A part of the
autoresonant orbit in the phase plane. Knowing the Hamiltonian at
the 4 points, we can calculate the maximum and minimum deviations
of the action and phase. The parameters are the same as in Fig.
\ref{fiq-C3}. \label{fiq-C6}}
\end{figure}

Fig.~\ref{fiq-C6} shows a part of a typical autoresonant orbit in
the phase plane. For $\nu(t)=const.$  this orbit is determined by
the equation $H(I,\psi,\nu)=const.$, and it is closed. As in our
case $\nu(t)$ changes with time, the trajectory is not closed. To
calculate the maximum and minimum deviations of action and phase
we should know the values of the Hamiltonian at 4 points of the
orbit that we will call ``up", ``down", ``left", and ``right" in
the following.

\begin{figure}[h]
\vspace{-0.1 cm} \leftline{\epsfxsize=7.5 cm 
\epsffile{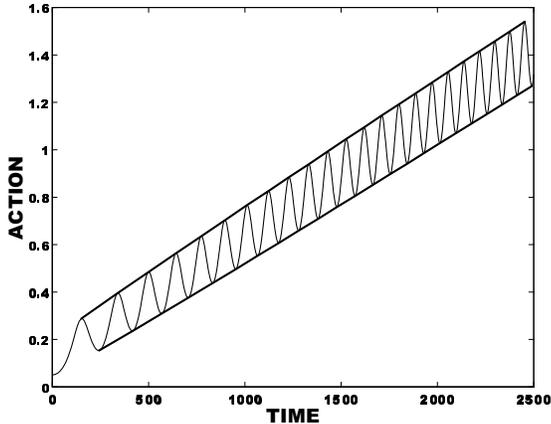}} \caption{The maximum and
minimum deviations of the action, calculated from Eq.~(\ref{C20})
(thick line) and from numerical solution (thin line). The
parameters are the same as in Fig. \ref{fiq-C3}. \label{fiq-C7}}
\end{figure}

\begin{figure}[h]
\vspace{-0.5 cm} \leftline{\epsfxsize=7.5 cm 
\epsffile{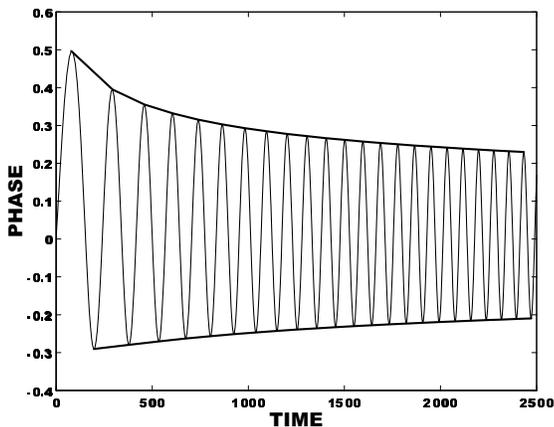}} \caption{The maximum and
minimum deviations of the phase, calculated from Eq.~(\ref{C21})
(thick line) and from numerical solution (thin line). The
parameters are the same as in Fig. \ref{fiq-C3}. \label{fiq-C8}}
\end{figure}

Knowing the values of the Hamiltonian at these 4 points, we
calculate $I_{max,min}$ from Eq.~(\ref{C20}) and $\psi_{max,min}$
from Eq.~(\ref{C21}). Figs.~(\ref{fiq-C7}) and~(\ref{fiq-C8}) show
these deviations for action and phase correspondingly, and the
values of $I$ and $\psi$, found from numerical solution. The
theoretical and numerical results show an excellent
agreement.

Now we are prepared to calculate the adiabatic invariant
$J(H,\nu(t))$. Its (approximate) constancy in time allows one, in
principle, to find the Hamiltonian $H(t)$ at any time $t$, in
particular at the points of the maximum and minimum action and
phase deviations (see Fig.~\ref{fiq-C6}).

It is convenient to rewrite the adiabatic invariant in the following
form:
\begin{equation}
J=\frac{1}{2\pi}\oint\psi dI. \label{C22}
\end{equation}

Using Eq.~(\ref{C1}), we can find $\psi=\psi(H,I,\alpha(t))$:
\begin{equation}
\psi=\pm \frac{1}{2}\arccos\left(\frac{8H+3\beta I}{2\epsilon I}
-\alpha\right)\,, \label{C23}
\end{equation}
so that Eq. (\ref{C22}) becomes:
\begin{equation}
J=\frac{1}{2\pi}\int^{I_{max}}_{I_{min}}\arccos\left(\frac{8H+3\beta
I}{2\epsilon I} -\alpha\right) dI, \label{C24}
\end{equation}
where $I_{max}$ and $I_{min}$ are given by Eq.~(\ref{C20}). Notice
that $H(t)$ and $\alpha(t)$ should be treated as constants under
the integral (\ref{C24}), see Refs.
\cite{Landau,Sagdeev,Goldstein}. This integral can be expressed in
terms of elliptic integrals (see Appendix B for details). For
definiteness, we used the values of $H(t)$ and $\alpha(t)$ in the
``up'' points, see Fig.~\ref{fiq-C6}. We checked numerically that
the adiabatic invariant $J(H(t),\alpha(t))$ is constant in our example
within 0.12 per cent.

Now we calculate the period of action and phase oscillations. From
the first equation of system ~(\ref{C2}) we have:
\begin{equation}
T=2\int^{I_{max}}_{I_{min}}\frac{dI}{(\epsilon I/2)\sin 2\psi},
\label{C26}
\end{equation}
where $I_{max}$ and $I_{min}$ are given by Eq.~(\ref{C20}), while
$\psi=\psi(I)$ is defined by~(\ref{C23}).

Using Eq.~(\ref{C1}), we obtain after some algebra:
\begin{equation}
T=\frac{8}{3\beta}\int^{I_{max}}_{I_{min}}\frac{dI}{G(I)^{1/2}},
\label{C27}
\end{equation}
where $G(I)$ is given in Appendix B, Eq. (\ref{C-G}). Again, we
treat $H(t)$ and $\alpha(t)$ as constants under the
integral~(\ref{C27}), and take their values in the ``right''
points, see Fig.~\ref{fiq-C6}. The final result is:

\begin{equation}
T=C_{2}K(C_{3}), \label{period}
\end{equation}
where $C_{2}=4(2/3\beta H\epsilon^2)^{1/4}$ and
$$C_{3}=\frac{1}{2}-\frac{C_{2}^2}{16}\left[\frac{3\beta
H}{2}+\frac{\epsilon^2}{16}\left(1-\alpha^2\right)\right]\,.$$

\begin{figure}[h]
\vspace{-0.0 cm} \rightline{\epsfxsize=7.5 cm 
\epsffile{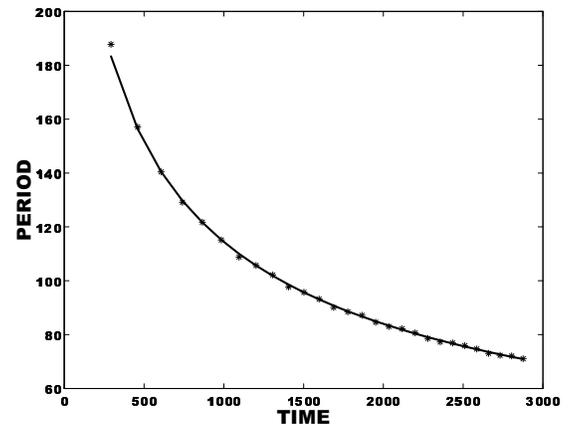}} \caption{The period $T$ of the phase (action)
oscillations obtained from Eq.~(\ref{period}) (solid line), and
from numerical solution (asterisks).
The parameters are the same as
in Fig. \ref{fiq-C3}.
\label{fiq-C10}}
\end{figure}

Figure \ref{fiq-C10} shows the period $T$ of the phase and action
oscillations versus time obtained analytically and from numerical solution.
This completes our consideration of the parametric autoresonance without
dissipation.

\subsection{Role of dissipation}

\begin{figure}[h]
\vspace{0 cm} \leftline{\epsfxsize=7.5 cm 
\epsffile{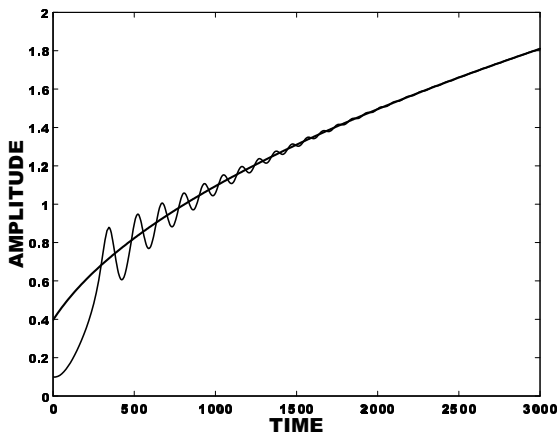}}

\caption{Parametric autoresonance with
dissipation: the time dependence of the amplitude
of oscillations, obtained from numerical solution of
Eqs. (\ref{B8}), and the amplitude trend $a_*(t)$, predicted by
Eq. (\ref{C29}). The parameters are $\mu=6.5\cdot 10^{-5},
\epsilon=0.04, \delta=0, \gamma=0.002$ and $\beta=1/6$.
\label{fiq-C13}}
\end{figure}

Now we very briefly consider the role of dissipation in the
parametric autoresonance. Consider the
averaged equations~(\ref{B8}) and assume that the detuning is zero.
The non-trivial
quasi-fixed point exists when the dissipation is not too strong:
$\gamma < \epsilon/4$, and it is given by
\begin{eqnarray}
a_{*}&=&\left(\frac{2\epsilon}{3\beta}\right)^{1/2}\left[\alpha (t) +
\left(1-\frac{16\gamma^2}{\epsilon^2}\right)^{1/2}\right]^{1/2},\nonumber
\\
\psi_{*}&=&\frac{1}{2}\arcsin\left(\frac{4\gamma}{\epsilon}+
\frac{2k}{\alpha (t) +(1-16\gamma^2/\epsilon^2)^{1/2}}\right)\,.
\label{C29}
\end{eqnarray}
Again, we assume $k\ll 1$. This quasi-fixed point describes the
slow trend in the dissipative
case.
As we see numerically, fast oscillations around
the trend,  $\delta a=a-a_{*}$ and $\delta\psi=\psi-\psi_{*}$
decay with time. Therefore,
one can expect that the $a(t)$ will approach, at
sufficiently large times, the trend $a_{*}(t)$.
Fig.~\ref{fiq-C13} shows the time dependence of the
amplitude, found by solving numerically the system of averaged
equations~(\ref{B8}), and the amplitude trend
from~(\ref{C29}). We can see that indeed
the amplitude $a(t)$ approaches the trend $a_{*}(t)$ at large times.

Therefore, a small amount of dissipation enhances the stability of
the parametric autoresonance excitation scheme. A similar result
for the externally-driven autoresonance was previously known \cite{Yariv2}.

\section{CONCLUSIONS}

We have investigated, analytically and numerically, a combined
action of two mechanisms of resonant excitation of
nonlinear oscillating systems: parametric resonance and
autoresonance. We have shown that parametric autoresonance represents a
robust and efficient method of excitation of nonlinear oscillating
systems. Parametric autoresonance can be extended for the
excitation of nonlinear {\it waves}. We expect that
parametric autoresonance will find applications in different
fields of physics.

\section*{ACKNOWLEDGEMENTS}
This research was supported by the Israel Science Foundation,
founded
by the Israel Academy of Sciences and Humanities.
\appendix

\section{CALCULATION OF PHASE AND ACTION DEVIATIONS BY THE WKB-METHOD}

Changing the variables from time $t$ to $\alpha$, we can rewrite
Eq.~(\ref{C12}) in the following form:
\begin{equation}
\delta\psi^{\prime\prime}+
\left(\frac{\alpha(t)+1}{4k^2}\right)\delta\psi=0\,, \label{AP31}
\end{equation}
where $^{\prime\prime}$ denotes the second derivative with respect to $\alpha$.
Solving this equation by the
WKB-method \cite{Lichtenberg}, we obtain
for $\delta\psi$:
\begin{equation}
\delta\psi(t)=\frac{(2kC)^{1/2}}{(\alpha+1)^{1/4}}\cos\left(\Omega_{0}+\frac{(\alpha(t)+1)^{3/2}-1}{3k}\right),
\label{AP36}
\end{equation}
where $\Omega_{0}$ and $C$ are constants to be found later.
Now we obtain the solution for $\delta I$.
Substituting~(\ref{AP36}) into the second equation of the
system~(\ref{C11}), we obtain in the same order of accuracy:
\begin{eqnarray}
\delta
I(t)&=&\frac{2\epsilon}{3\beta}\left(2kC\right)^{1/2}\left(\alpha+1\right)^{1/4}\nonumber
\\
& \times &
\sin\left(\Omega_{0}+\frac{(\alpha(t)+1)^{3/2}-1}{3k}\right).
\label{AP37}
\end{eqnarray}
The constant $C$ can be expressed through the adiabatic invariant
$J$, given by~(\ref{C7}). From Eqs.~(\ref{AP36}) and~(\ref{AP37})
we have: $$
2kC=\left(\frac{3\beta}{2\epsilon}\right)^2\frac{1}{(\alpha+1)^{1/2}}\left(\delta
I\right)^2+\left(\alpha+1\right)^{1/2}(\delta\psi)^2.$$ Comparing
it with~(\ref{C7}) we find: $ C\simeq 3\beta J/2k\epsilon.$
Substituting this value into Eqs.~(\ref{AP36}) and~(\ref{AP37}) we
obtain the final expressions~(\ref{C15}) and (\ref{C19}) for
$\delta\psi(t)$ and $\delta I(t)$.

\section{CALCULATION OF THE ADIABATIC INVARIANT}

After integration by parts and some algebra, using
Eqs.~(\ref{C1}) and ~(\ref{C20}), we obtain the following
expression for the
adiabatic invariant:
\begin{equation}
J=\frac{1}{2\pi}\int^{I_{max}}_{I_{min}}\left(\frac{I^2-
\frac{8H}{3\beta}}{G(I)^{1/2}}\right)
dI, \label{C25}
\end{equation}
where
\begin{equation}
G(I)=\left(I_{max}-I\right)\left(I-I_{min}\right)\left[\left(I+\frac{\epsilon(1-\alpha)}{3\beta}\right)^2-\frac{16D}{9\beta^2}\right],
\label{C-G}
\end{equation}
and we assume $D=(\epsilon^2/16)(1-\alpha)^2-3\beta H/2<0.$
Calculation of this integral employs several changes of
variable shown in the best way by Fikhtengolts
\cite{Fikhtengolts}. Using the reduction formulas
\cite{Abramowitz}, we arrive at:
\begin{eqnarray}
J=C_{1}\left[\frac{1+mm'}{(1-m)^2(1+m')}\Pi\left(\frac{m}{m-1}\backslash
k^2\right)\right.\nonumber
\\
\left.-\frac{1}{1-m}K\left(k^2\right)+\frac{m+m'}{(1-m)(1+m')}E\left(k^2\right)\right],
\label{invariant}
\end{eqnarray}
where
$$m=\frac{(\epsilon/3\beta)(1+\alpha)-(8H/3\beta)^{1/2}}{(\epsilon/3\beta)(1+\alpha)+(8H/3\beta)^{1/2}}>0,$$
$$m'=\frac{(\epsilon/3\beta)(1-\alpha)+(8H/3\beta)^{1/2}}{-(\epsilon/3\beta)(1-\alpha)+(8H/3\beta)^{1/2}}>0.$$
$$k^2=\frac{m}{m+m'},\indent
C_{1}=c\cdot\frac{64H}{3\beta(m+m')^{1/2}},$$
and
\begin{eqnarray}
c&=&\frac{1}{2\pi}\left[\frac{\epsilon}{3\beta}\left(1+\alpha\right)+\left(\frac{8H}{3\beta}\right)^{1/2}\right]^{-1/2}\nonumber
\\
& \times &
\left[-\frac{\epsilon}{3\beta}\left(1-\alpha\right)+\left(\frac{8H}{3\beta}\right)^{1/2}\right]^{-1/2}\,.\nonumber
\end{eqnarray}
Here $K$, $E$ and $\Pi$ are the complete elliptic integrals of the
first, second and third kind, respectively.

\end{document}